# Superstars in politics: the role of the media in the rise and success of Junichiro Koizumi

Eiji Yamamura[1],[2]
Fabio Sabatini[3]

Abstract

This paper explores the role of mass media in people's perceptions of charismatic leaders, focusing on the case of Junichiro Koizumi, Prime Minister of Japan from 2001 to 2006. Using survey data collected immediately after Koizumi's 2005 landslide electoral victory, this study empirically assesses the influence of television (TV) and newspapers on individuals' support for Koizumi and for the most distinctive policy action he announced during his campaign—the privatization of the postal service.

The major findings are: (1) the frequency of exposure to mass media is positively related to the support for Koizumi but not for his principal policy and (2) a significant impact of TV is only observed among women. The habit of reading newspapers only slightly correlates with men's support for Koizumi. Our study's results suggest that compared to a political platform, charisma and attractiveness wield a greater influence on TV watchers of the opposite sex.

Television apparently works as a powerful amplifier of leaders' appealing attributes. The resulting superstar effect may allow a charismatic candidate to win an election, even though his main agenda item (i.e., postal privatization) is strongly opposed by special interest groups and members of the ruling party.

[1] Department of Economics, Seinan Gakuin University, Japan, Institute of Social and Economic Research (ISER) Osaka University, Japan, and Laboratory for Comparative Social Research, National Research University Higher School of Economics, Russia.

[2] Corresponding Author. Postal Address: Department of Economics, Seinan Gakuin University, 6-2-92 Sawaraku Nishijin, Fukuoka 814–8511, Japan.

[3] Department of Economics and Law, Sapienza University of Rome, and Laboratory for Comparative Social Research, National Research University Higher School of Economics, Russia.

## 1. Introduction

Popularity is one of the key determinants of the increase in market demand. For example, in the entertainment or sports market, the demand for superstars is far greater than that for less popular people, resulting in a substantial gap of earnings between superstars and others. The theory of superstars suggests that the celebrity effect stems from imperfect substitutability and the ability to attract an audience and stimulate huge amounts of consumption in large markets (Rosen 1981). Mass media such as television (TV) and newspapers have the power to increase the market size and generate a large volume of transactions that may specifically benefit superstars. Superstar effects on earnings have been found in several sectors of the labor market, from sports (Lucifora and Simmon 2003) to gastronomy (Ehrmann et al. 2009).

We extend the analysis to politics, arguing that mass media play a key role in creating superstar effects that may easily determine electoral outcomes. Contemporary elections are built on a visual foundation (Grabe and Bucy 2009). Candidates must appear telegenic enough to attract closer scrutiny by their audience. Electoral campaigns are devoted to strengthening candidates' personal appeal, often to a larger extent than proposals for policy initiatives and programs. The rise of TV as a political force has created incentives for political leaders to construct their image through deliberate strategies designed to promote intimacy with voters and to highlight engaging personal attributes.[4] Since in the market of politics, the superstar effect entails enduring electoral support, it is important to explore how exposure to mass media influences voters' support for political actors and their economic agenda.

The previous literature analyzed the relationship between exposure to media and specific political outcomes. Based on US historical data, Strömberg (2004) found that larger relief funds were allocated to counties where higher shares of households listened to the radio. Gerber and colleagues (2009) used a field experiment conducted in Washington to analyze the effect of newspapers' slant on readers' political knowledge, stated opinions, and voter turnout. The authors found that reading a liberal newspaper significantly increased support for the Democratic candidate for governor. Another study reported that the introduction of the conservative Fox News Channel in 20% of US towns led its viewers to vote for conservative Republican candidates (Della Vigna and Kaplan 2007). Spanish-language news programs on TV also contributed to a substantial boost in voter turnout (Oberholzer-Gee and Waldfogel 2009). Researchers have increasingly investigated the mass media and their political influence in other countries such as Italy (Durante and Knight 2012; Sabatini 2012), Russia (White et al. 2005; Enikolopov et al. 2011), Mexico (Lawson and McCann 2005), India (Olken 2009),[5] and Muslim countries (Gentzkow and Shapiro 2004). These studies have revealed mass media's critical role in establishing trends in public opinion.

Since the end of the 1980s, mass media, especially TV, has wielded a crucial influence on politics in Japan. According to Krauss and Nyblade (2005), this situation has led the Japanese public to "consume" politics. This trend was considerably accelerated by the rise of Junichiro Koizumi, Prime Minister (PM) from 2001 to 2006. Koizumi has been regarded as the most charismatic political leader in Japan over the last 25 years. He is generally considered likeable and attractive; the *New York Times* described Koizumi as Japan's best-known Elvis impersonator (Stolberg and Dewan 2006). Koizumi also shares similarities with Ronald Reagan in that they both communicated with voters using a performance-based style, and they advanced intense programs of neoliberal reforms aimed to reduce the size of government, deregulate industries, and increase market competition. Koizumi cleverly exploited his style, attractive appearance, and TV performances to attract media attention and gain the support of the majority of voters, despite the fierce opposition to his reforms in Japanese society and in the national political debate (Imai 2003; Ohtake 2003).

---

[4]It was found that better-looking people earn more than average-looking people (Hamermesh and Biddle 1994; Biddle and Hamermesh 1998). To take the example of sports, attractiveness, as measured by facial symmetry, leads to greater rewards even after controlling for player performance (Berri et al. 2011).

[5]Research related to media and economic issues has also been conducted based on cross-country data (Djankov 2003; Connolly and Hargreaves Heap 2007; Petrova 2008).

We use individual-level data drawn from the "Social and Political Consciousness Survey in 21st Century Japan" (GLOPE) conducted in 2005 to assess the determinants of public consensus in support of Koizumi's administration, with a focus on the role of mass media. More specifically, we examine how the frequency of watching TV and reading newspapers influenced public support for Koizumi and his principal policy. The GLOPE survey was conducted in Japan immediately after the 2005 election to determine citizens' perceptions about politics during an election period. In his electoral campaign, Koizumi deliberately and strongly drew the electorate's attention to his program of postal service privatization, and he won by a landslide.[6]

This paper makes several contributions to the relevant literature. First, this study empirically assesses whether the superstar effect may also hold true in the political market. It disentangles the possibly competing roles of different mass media in popularizing a charismatic leader's image and attracting political consensus. The GLOPE data also allow us to illustrate how citizens' votes for Koizumi were virtually disconnected from their support for the candidate's program of public sector reforms. Arguably, the case of a celebrity in the Japanese political arena is useful in analyzing how the superstar effect can secure electoral victory, despite the opposition of special interest groups.

The rest of the paper is organized as follows. Section 2 provides the political background and overview of Koizumi's policies and strategies of communication. Subsection 2.3 presents two testable hypotheses about citizens' views on Koizumi's administration and platform. Section 3 describes the dataset and provides a simple econometric framework to examine the hypotheses. Estimation results are reported in section 4. The final section offers concluding remarks.

## 2. The Setting and Hypotheses

*2.1. Political Background*

The Liberal Democratic Party (LDP) has been Japan's dominant party for more than 50 years after it was founded in 1955 (apart from two short periods in 1993–1994 and 2009–2012). It is a major conservative party, whose economic policies have historically been shaped around the goals of free trade, market competition, and export-based economic growth.[7] Junichiro Koizumi is known as the LDP's most popular PM. At the helm from 2001 to 2006, he was the fifth longest-serving PM in Japanese history. Table 1 shows that in the 1989–2012 period, Koizumi's government lasted for 60 months, which was longer than any other recent administration (the second longest one was Hashimoto's 30-month administration, half the length of Koizumi's). In 2005, Koizumi led the LDP to win one of the largest parliamentary majorities in modern Japanese history. According to opinion polls, public support for Koizumi peaked at 85%, the highest rating since 1989. The lowest support rate during the Koizumi administration was 39%, with only those of the Hosokawa and Hata governments being higher. Considering that the Hosokawa and Hata governments only lasted 8 and 2 months, respectively, the lowest rating received by Koizumi was in fact quite high, compared with that of any other administration.[8] To understand the reason why Koizumi garnered such high levels of support, it is necessary to discuss briefly the political and economic background of Japan's post-World War II period.

The LDP is not a cohesive organization but a conglomeration of competitive factions (*habatsu*), which engage in bitter infighting. *Habatsu* play a fundamental role in every area of LDP political activity, including candidates' nominations, the acquisition of funds, and the allocation of party and government posts (Köllner 2006). According to Köllner (2006), "The LDP's institutionalized factional system served as an effective functional equivalent of formalized procedures and norms of party management … Informal rules on how party and cabinet posts were to be allocated made the political careers of LDP Diet members more foreseeable and helped to reduce uncertainty" (p. 248). For a long time during the LDP's ruling period, various factions formed links with special interest groups and industries, which are considered to play key roles in maintaining the LDP's electoral dominance. In fact, in Japan, politicians have been suspected of increasing the expenditure allocated to public infrastructures, with the aim of attracting the support of the construction industry (Yamamura and Kondoh 2013). This case may explain why "larger amounts are spent on public works than in other countries, controlling for size and population" (Doi and Ihori 2009, p. 181). The LDP members of the Diet also form committees called "policy tribes" (*zoku-giin*), which serve to favor those interest groups that provided support in electoral campaigns.[9] In turn, thanks to the strong electoral support from these groups, the LDP maintained a one-party cabinet for 38 years during the 1955–1993 period.[10]

---

[6] No significant relationship was observed between the results of the 2005 election in Japan and happiness (Tsutsui et al. 2010).

[7] The LDP must not be confused with the now-defunct Liberal Party, which merged with the Democratic Party of Japan, the main opposition party, in November 2003.

[8] The Hosokawa and Hata regimes were the first non-LDP regimes after World War II. They only received support for a brief period in 1993–1994.

In the early 1990s, the LDP's political power and internal stability declined as a consequence of corruption scandals. Some influential LDP members left the organization and founded new parties that entered into coalitions with existing ones. Eventually, in August 1993, a seven-party coalition opposed to the LDP formed a new government that stayed in charge for 10 months. As a response, the LDP formed a new alliance with the Social Democratic Party. Since that time, the LDP has always been forced to share its governmental responsibilities with other parties to ensure a parliamentary majority. From the first half of the 1990s, Japan experienced long-term economic stagnation, which further undermined the LDP's predominance. The political influence of specific interest groups (which mostly included representatives of rural areas) declined, and the mobilization of unaffiliated and undecided voters (mainly living in urban areas) became increasingly important. At the beginning of the new millennium, the LDP thus faced a turning point.

*2.2. Koizumi's Victory in the 2005 Election*

In 2001, Koizumi was elected president of the governing LDP. He cleverly exploited his flamboyant and colorful political style to gain the vote of prefectural party organizations, whose rank-and-file members were longing for a plain-spoken leader with charisma and a vision for a new Japan. Koizumi pledged to scrap political conventions, appointing cabinet members based on expertise and merit rather than on their membership in *habatsu* (Strom 2001). With his provocative, conflict-oriented political style, Koizumi rapidly attracted massive media attention. From the beginning of his political career, he often appeared in traditional mass media, such as the national newspapers and hard news on TV. However, when his popularity increased, he extended his coverage by participating in soft news, TV gossip shows and talk shows. Koizumi's interest in sports also made him an interview object of sports magazines.

At that time, Koizumi belonged to the Fukuda group, a faction that became antagonistic to the clientelistic politics that the mainstream LDP represented. Hence, he did not enjoy any particular benefit from the LDP's institutionalized factional system. To overcome his isolation within the LDP, Koizumi had a strong incentive to promote political reforms and use the media to gain external consensus. "Koizumi managed to capture the imagination of the Japanese public by means of snappy slogans and dramatizing politics. What set Koizumi apart from most former LDP leaders was that he appealed directly to [the] public" (Köllner 2006, p. 251). He presented himself as the spokesman of the people in a struggle with the Japanese elite, and he promised the end of preferential treatments for the groups and industries that until then sponsored the party factions (Lindgren 2012). In his political message, Koizumi insisted on the need to destroy the unhealthy relationships among politicians, bureaucrats, and entities with vested interests as the main challenge to rescue the Japanese economy. For this purpose, he repeatedly declared being ready to restrain his own party and its factions, as well as the industries that so far provided political support to the LDP.

---

[9]The relationships among politicians, bureaucrats, and private companies are sometimes described as an "iron triangle" (Sakakibara 2003).

[10]During the period, the rival party had been the Social Democratic Party of Japan, which is supported mainly by labor unions. However, the Social Democratic Party had never come to power until the 1990s.

As part of this strategy, he proposed an intensive program of neoliberal reforms, including deregulation of industries (especially the "protected" ones), reduction of government size and public spending, privatization of public corporations, and introduction of corporate models of management in public welfare institutions. These measures not only had the objective of revitalizing the Japanese economy; during his term as PM, his political message also aimed to disconnect the ties among politicians, bureaucrats, and protected industries—to let prices, profits, and employment levels be governed by market forces, instead of being the result of a bargaining process between the political establishment and business organizations. The LDP supporters were roughly divided into those who would benefit from the reforms and those who would be hurt by them. "Koizumi came to power via a revolt by the LDP's grassroots urban machine against the more rural-oriented party leadership" (Katz 2001, p. 38). He "demonstrated an uncanny ability to instrumentalize the media, both old and new. Electoral cooperation with the new Komei party progressed quite smoothly under Koizumi, contributing to the parliamentary dominance of the LDP" (Köllner 2006, p. 254).

In 2005, the privatization of the postal service (considered as Koizumi's principal policy) was strongly contradicted, not only by members of the opposition party, but also by many LDP members. Inevitably, the bill to privatize the postal service was rejected in the House of Councilors in August 2005. The rebellion of his own party members essentially prevented Koizumi's primary policy from being realized. In response, Koizumi dissolved the House of Representatives (*Shugi-in*) and called for a general election to seek the public's opinion on the bill. The privatization of the postal service thus became a key issue of the electoral campaign. In other words, Koizumi framed the election as a referendum on the privatization of the postal service.

The *Shugi-in* has 480 members who are elected for a four-year term. Among them, 300 members are elected in single-seat constituencies and 180 members by proportional representation in 11 districts. In each constituency, only one candidate can be elected. To avoid internal conflicts, each party usually nominates a lone candidate for a constituency. However, in the 2005 election, Koizumi broke this rule and recruited so-called "assassin" candidates to stand against the LDP members (representing some constituencies) who did not endorse the bill of postal privatization. By this means, Koizumi brought further drama to the vote. During the campaign, he made massive use of the mass media, especially TV, to describe himself as the hero of the reform and his opponents within the LDP as "resistance forces" that were "stuck in the past" (Köllner 2006). He clearly turned the electoral competition into a spectacular showdown between himself and his rivals. Overall, the electoral campaign was transformed into a sort of TV game show, encouraging voters to enjoy politics as they did sports or movies (Ohtake 2003). Consequently, the TV audience remarkably increased, leading TV broadcasts to treat the campaign as a form of political entertainment.

On one occasion, Koizumi danced in his office with the American actor Richard Gere (who has been said to resemble him) (Van Gelder 2005). Political analysts argued that since this event was widely broadcasted on TV, Gere's Japanese fans would then like Koizumi, even if they were indifferent to politics. The *New York Times* also reported an impressive episode in the US, emblematically illustrating the PM's colorful political style. In 2006, Koizumi visited Elvis Presley's mansion with the then President George W. Bush. Koizumi shares the same birthday (January 8) as that of Elvis; their similar hairstyles have also been noted. Koizumi rendered an Elvis Presley impersonation during his tour of the singer's home. When Elvis' former wife pointed out a pair of oversized, gold-rimmed sunglasses once worn by the King of Rock and Roll, the PM eagerly donned them, thrusting his hips and arms forward in imitation of a classic Elvis move. He later threw his arm around Elvis' daughter, belting out some Elvis lyrics, "hold me close, hold me tight" (Stolberg and Dewan 2006).

Koizumi's strategy apparently managed to catch the attention of those who were usually uninterested in politics. It enhanced the common people's involvement in the political debate; as a result, undecided and less informed voters massively turned to the LDP in the election (Köllner 2006). Mass media played a fundamental role in inducing undecided or unaffiliated urban voters to support Koizumi's principal policy. Where the influence of special interest groups and factions declined, "the personalization of the role [of the PM] is increasingly important to voters … Skillful and attractive prime ministers will gain popularity and better results for their party; unskillful and unattractive ones will find their terms quite short" (Krauss 2002, p. 12).

In Japan, newspapers can be roughly divided into two categories. The national newspapers (such as *Asahi*, *Yomiuri*, and *Mainichi*) are distributed throughout Japan, and minor local newspapers also abound. Generally, the readers of the former are far greater in number than those of the latter. For instance, in 2010, the circulation of *Yomiuri* was 10 million, and that of *Asahi* was 7.9 million. To put it in context, compare it with the circulation of the *Wall Street Journal* and *USA Today* in the US at 2.0 million and 1.8 million, respectively.[11] These differences in numbers show the newspapers' wide circulation in Japan, compared with those of other developed countries. In 2005, 99.3% of Japanese households had TV sets.[12] Therefore, considering the combined newspaper readership and TV viewership during the Koizumi period, the Japanese people were likely highly exposed to both media forms.

*2.3. Empirical Hypotheses*

Gentzkow (2006) described the introduction of TV in the 1950s as a dramatic improvement of previous entertainment technologies. Accordingly, the price of political information decreased, with an even greater reduction in entertainment cost. Consumers massively responded by moving away from political news toward new forms of entertainment. However, Koizumi's ability to dramatize politics allowed TV broadcasts to incorporate political news into entertainment programs. Political news and entertainment began to be broadcasted in the same programs. According to Dyck and colleagues (2013), "By collecting news and combining it with entertainment, media are able to inform passive voters about regulation and other public policy issues, acting as a [partial] counterbalance to small but well-organized groups" (p. 521).

However, mass media also draw citizens' attention to the personal characteristics of political leaders rather than to policy issues. This process reduces the role and visibility of special interest groups, as well as the importance of *habatsu*. A similar example can be observed in the relationship between a movie and its leading actor. Fans of the leading actor often watch the film just because they want to see their idol on screen, even if they are not interested in the movie itself. This pattern also holds true for prominent actors competing in elections, as in Ronald Reagan's case in American politics. To test Dyck and colleagues' (2013) argument about the role of mass media in shifting public opinion's attention from specific policy issues to the detriment of special interest groups, it is essential to empirically disentangle Koizumi's popularity from that of his policy proposals that raised the opposition of special interest groups. Therefore, we propose the following two hypotheses:

*Hypothesis 1: As mass media contribute to spreading detailed information about policy issues, exposure to mass media may lead voters to support specific policies.*

---

[11] Even though since the late 1990s, the circulation of Yomiuri has been growing, while that of local newspapers declined (George and Waldfogel 2006), at of the beginning of 2010, the circulation of the *New York Times* was about 0.9 million. . The data source of circulation in 2010 is "Circulation Data For National Newspapers", which is available on the website of The International Federation of Audit Bureau of Circulations (IFABC):
[12] The data source is "Consumer Confidence Survey", which is available on the website of the Cabinet Office, Government of Japan:
http://www.esri.cao.go.jp/jp/stat/shouhi/shouhi.html (Accessed on May 7, 2014).

*Hypothesis 2: As mass media highlight the appealing personal attributes of charismatic leaders, exposure to mass media may lead voters to support a leader even if they are not interested in his or her principal policy.*

We argue that the case of Koizumi's success is particularly appropriate to test *hypotheses 1* and *2*, due to the Japanese PM's strikingly engaging characteristics and special interest groups' fierce opposition to his platform.

### 3. Estimated Model and Estimation Results

*3.1. Data*

The GLOPE survey was purposefully conducted throughout Japan after the 2005 election to systematically investigate voters' political opinions at the end of the electoral campaign. The respondents were adults belonging to various generations. Sample points could be divided into cities (considered as urban areas) and towns or villages (considered as rural areas). Three thousand adults were invited to participate in the survey, with a stratified, two-stage random sampling. The data included detailed information on how the Koizumi administration and his principal policy (postal service privatization) were perceived, as well as the respondents' home ownership, years of schooling, demographic characteristics (age and gender), and household income.[13]

Information on the sample used in this research is reported in Table 2. The mean value of *KOIZUMI* (Koizumi supporter dummy) is 0.62, meaning that 62% of the people supported Koizumi's government. This figure is consistent with the election results. The mean value of *PRIVAT* (postal privatization supporter dummy) is 0.42, suggesting that 42% of the people supported the privatization of Japan's postal service. This 20% difference implies that one-third of Koizumi's supporters were attracted by his appealing personality but did not favour his main policy. It seems possible that people voted for the LDP and its candidates even though they paid little attention to the issues debated during the electoral campaign. Figure 1(a) indicates no relation between the frequency of watching TV and the men's support for the Koizumi administration. However, Figure 1(b) suggests that the more frequently the women watched TV, the more likely they would support Koizumi's government. Neither Figure 2(a) nor Figure 2(b) reveals an association between the frequency of watching TV and the support for the privatization of postal service, regardless of the respondents' gender. These results imply that TV programs may have contributed to increasing the number of women who supported Koizumi but not his principal policy (a central issue in the general election).

On the other hand, the frequency of reading newspapers is only weakly related to the men's support for Koizumi and not at all for his policy.[14] Overall, the research results are consistent with *hypothesis 2* but not with *hypothesis 1*. To further scrutinize these findings, regression estimations were conducted, and the results are discussed in section 4.

*3.2. Empirical Model*

To test *hypotheses 1* and *2*, the estimated function takes the following form:

$Y_i = \alpha_0 + \alpha_1 TV_i + \alpha_2 NEWS_i + \alpha_3 Ln(INCOM)_i + \alpha_4 Ln(EDU)_i + \alpha_5 Ln(AGE)_i + \alpha_5 SPOUS_i + \alpha_6 CHILD_i + \alpha_7 UNEMP_i + \alpha_8 KNOW_i + \alpha_9 GOVPT1_i + \alpha_{10} GOVPT2_i + \alpha_{11} VILLAG_i + \alpha_{12} MAN$

---

[13] Data for this secondary analysis were sourced from the "Social and Political Consciousness Survey in 21st Century Japan (GLOPE 2005)". Data were collated by the Waseda University Centre of Excellence Program for the 21st Century: Constructing Open Political-Economic Systems (21 COE-GLOPE). The research was subcontracted to Chuo Chosa-Sha. Data were provided by the Social Science Japan Data Archive, Information Center for Social Science Research on Japan, Institute of Social Science, The University of Tokyo.

[14] The figures may be requested from the corresponding author.

$_i + u_i$ ,

where $Y_i$ represents the dependent variable for an individual $i$ such as *KOIZUMI* and *PRIVAT*. The error term is represented by $u_i$. Because *KOIZUMI* and *PRIVAT* are dummy variables that are either 1 or 0, a probit model can be used. However, if a correlation exists between the residuals of different equations, then a bivariate probit model is more appropriate to conduct estimations of *KOIZUMI* and *PRIVAT* at the same time (Greene 2008). Furthermore, $\alpha'$ is the marginal effect, as reported in Tables 3, 4, and 5.[15]

The coefficients of *TV* and *NEWS* might be positive in columns (2), (4), and (6) of Tables 3-5, in case *hypothesis 1* would be supported. On the other hand, the coefficients of *TV* and *NEWS* might be positive in columns (1), (3), and (5), in case *hypothesis 2* would be supported. This approach would allow the comparison of the effects of the two forms of media. Our results suggest that newspapers could provide more detailed information about political issues than TV shows could. However, TV broadcasts of Koizumi's performances were more vivid and visual than newspaper articles and more effective in attracting stronger consensus with his government, despite the audience's lack of attention to the privatization policy announced during the electoral campaign.

To control for the respondents' socioeconomic conditions, the logs of household income level (*INCOM*), years of schooling (*EDU*), age (*AGE*), and work status (*UNEMP*) were included as independent variables. Generally, Koizumi's economic agenda relied on free trade and market competition as drivers to improve economic efficiency. The privatization of the postal service was intended as the first step of a broader program aimed at reshaping the public sector. In principle, the expansion of the private sector might offer highly educated people greater opportunities for higher earnings. Therefore, *EDU* was predicted to be positive. In contrast, unemployed workers were less likely to benefit from the restructuring of the public sector, so the coefficient of *UNEMP* was expected to be negative. *SPOUS* (dummy for being married) and *CHILD* (dummy for having children) were incorporated to capture the family structure. *KNOW* (a dummy equal to 1 if the interviewee could identify the Speaker of the House of Representatives in Japan) was included to capture the degree of the respondents' knowledge of Japan's political situation. As mentioned, Koizumi's government was supported by an alliance between the LDP and the new Komei party. The respondents' political standpoints were captured by *GOVPT1* (dummy for being an LDP

---

[15] In a similar methodological framework, Sabatini (2012) used an instrumental variables approach to control for endogeneity of trust in TV as a determinant of trust in former Italian PM Silvio Berlusconi. However, in the case addressed by Sabatini (2012), Berlusconi owned some major media outlets and actively controlled them, which created ambiguity about the interpretation of the relationship as a causal one. In fact, people who trusted Berlusconi might have a higher propensity to trust TV. In the Italian study, the key variable, trust in TV, is clearly endogenous due to the possibility of reverse causality. In the Japanese case discussed in the present paper, Koizumi never owned a media outlet, and there is a decent degree of pluralism in TV and newspaper ownership and contents. The Japan General Social Surveys (JGSS) have been conducted nationwide to collect individual-level data, more specifically, covering the 2000–2010 period during and after the Koizumi government. Based on the JGSS data, the average time spent watching TV was around 3.5 hours, which did not differ between the period of the Koizumi government and other periods. The JGSS data also indicated that the habit of reading newspapers did not change before, during, and after Koizumi's administration. These findings suggest that TV watching and newspaper reading remained stable, basically depending on lifestyles and other variables such as education (which are controlled for within the empirical analysis), rather than on Koizumi's performance or his principal policy. Therefore, the assumption that the key variables are endogenous does not hold true in the present paper.

Supplementary estimations are conducted to check this argument. For instrumental variables, Sabatini (2013) used proxies of social capital such as the quality of friendships and trust in the press. In the present study, we followed Sabatini's (2013) method and conducted an instrumental variable (IV) probit estimation where *TV* and *NEWS* were treated as endogenous variables and instrumented with the degree of community participation (regarded as a proxy for social capital). The results of the Wald test for the exogeneity of *TV* (and *NEWS*) were obtained. The test statistics were insignificant; thus, there is insufficient evidence in the sample to reject the null hypothesis about the absence of endogeneity. This finding suggests that there is no need to use instrumental variables because the endogeneity bias does not exist. The results of the IV probit model are similar to those reported in Tables 3–5. They are omitted for the sake of brevity and may be requested from the corresponding author.

supporter) and *GOVPT2* (dummy for being a supporter of the new Komei party). Reverse causality between *KOIZUMI* and *GOVPT1 (*and *GOVPT2)* was possible, resulting in estimation bias. However, the estimation results reported later did not change, even when *GOVPT1* and *GOVPT2* were excluded from the set of independent variables. Traditionally, the LDP received support from rural voters; however, the influence of urban residents (traditionally independent voters) also increased. As mentioned in section 2, Koizumi enjoyed greater support from urban residents than from their rural counterparts. Hence, the coefficient of *VILLAGE* was anticipated to be negative. Furthermore, the perception that females would find Koizumi more appealing led to a negative forecast for *MAN*.

**4. Estimation Results**

Table 3 reports the estimation results based on the whole sample. Tables 4 and 5 present the results for men and women, respectively. Tables 3–5 used a bivariate probit model; thus, the determinants of *KOIZUMI* and *PRIVAT* were jointly estimated. The pairs of joint estimations are reported in columns (1) and (2), (3) and (4), and (5) and (6). The correlation coefficient for *TV* and *NEWS* is 0.42 and statistically significant. Hence, to deal with the possibility of such collinearity, we also report alternative models where either *TV* or *NEWS* is omitted from the set of independent variables. Therefore, two sets of results for *KOIZUMI* and *PRIVAT* are presented in columns (3) and (4) and columns (5) and (6).

As expected, the coefficients of the key variables (*TV* and *NEWS*) are positive in columns (1), (3), and (5) of Table 3 when *KOIZUMI* is the dependent variable. *TV* is statistically significant in columns (1) and (3), and *NEWS* is statistically significant in column (5) but not in (1). Owing to the correlation between *TV* and *NEWS*, their standard errors increase in column (1), which might lead to insignificant values for *NEWS*. In contrast, the significance of *TV* does not change, suggesting robust results. Additionally, the size of the marginal effect of *TV* is larger than that of *NEWS* when *KOIZUMI* is the dependent variable. Overall, the effect of the frequency of watching TV on the support for Koizumi's government is more significant than that of reading newspapers.

Regarding the results of *PRIVAT*, contrary to predictions, *TV* is negative in columns (2) and (4) of Table 3. As predicted, *NEWS* is positive in columns (1) and (6). Neither *TV* nor *NEWS* is statistically significant. These findings suggest that exposure to mass media increases the voters' support for Koizumi. In contrast, *TV* and *NEWS* do not influence the voters' advocacy of Koizumi's principal policy, despite this being a key issue in the 2005 electoral campaign. Hence, these results confirm *hypothesis 2* but not *hypothesis 1*.

Concerning the control variables presented in Table 3, *INCOM* is positive in all columns. Furthermore, *INCOM* is statistically significant in the estimation of *KOIZUMI*, but not in that of *PRIVAT*. Respondents with higher incomes apparently are more likely to support Koizumi. However, income level is not correlated with the support for the privatization of the postal service. As shown in columns (1), (3), and (5), the coefficient of *EDU* is negative but statistically insignificant in the estimation of *KOIZUMI*. As reported in columns (2), (4), and (6), *EDU* is positive and statistically significant in the estimation of *PRIVAT*. Education seems to have conflicting effects on the support for Koizumi and that for his privatization program. On one hand, more educated voters are likely to perceive the reshaping of the public sector as a source of opportunities for the private sector that may benefit highly skilled workers. On the other hand, higher levels of education seem to make voters less sensitive to the appeal of leaders' personal characteristics.

These results are generally consistent with Sabatini's (2012) findings, which identified trust in TV as a strong predictor of trust in the Italian PM Silvio Berlusconi in 2011, thus suggesting his media empire's key role in amplifying his appealing personal attributes, thereby fundamentally contributing to building political consensus.

As reported in columns (1), (3), and (5) of Table 3, the coefficient of *AGE* is statistically significant and negative when the determinants of the support for Koizumi are estimated. These findings suggest that older people who share more traditional Japanese values are less influenced by Koizumi's performances, probably preferring more conventional ways to develop the political debate. With respect to family structure (*SPOUS* and *CHILD*), work status (*UNEMP*), and knowledge of politics (*KNOW*), their coefficients do not show statistical significance in any column. One possible explanation for these results is that the effects of work status and political knowledge are absorbed by education or income. Consistent with the prediction, *GOVPT1* is positive and statistically significant in all columns, suggesting the LDP supporters' tendency to support both Koizumi and his political platform. The results of *GOVPT2* are similar to those of *GOVPT1* although statistically insignificant in columns (2), (4), and (6). These findings reflect that the followers of the new Komei party (allied with the LDP) are inclined to support Koizumi rather than his specific policy. *VILLAGE* is negative in all columns and shows statistical significance in columns (1), (3), and (5). These results corroborate the argument that Koizumi was widely supported by undecided voters, mainly urban residents, but not by rural voters who were more likely to belong to some special interest groups in the agricultural sector (Miyake 1989).

Tables 4 and 5 show the gender differences in the results for the key variables. In Table 4, which only accounts for men's responses, the coefficients of *TV* and *NEWS* are positive in all columns but statistically insignificant, with the exception of *NEWS* in column (5). Thus, the frequency of watching TV is not associated with men's support for Koizumi's government and policy. The frequency of reading newspapers also has a weak effect on men's support. These results suggest the mass media's lack of influence on men's views about Koizumi.

Table 5 reports the results of the estimation conducted on the subsample of women. The coefficient of *TV* is positive and statistically significant in columns (1) and (3) and is negative and insignificant in columns (2) and (4). Furthermore, the statistical insignificance of *NEWS* in all the columns implies that the frequency of watching TV leads women to prefer Koizumi but not his political agenda. In contrast, the frequency of reading newspapers has no impact on women's advocacy for Koizumi's government and policy. It is interesting to observe that *VILLAGE* is not statistically significant although it is negative, indicating no difference between rural and urban women in their support for Koizumi. This finding is in line with the argument that Koizumi enjoyed widespread popularity and support in both urban and rural regions (Ohtake 2003, pp. 128–129).

Summarizing the results reported in Tables 3–5 leads to the conclusion that the combined estimation results proposed in this section are congruent with and reasonably support *hypothesis 2*, while *hypothesis 1* is not validated. Moreover, the images provided by the media, specifically on TV, do not influence the support from men but (probably due to TV's ability to highlight the PM's appealing personal traits) stimulate significant backing from women. Undecided female voters are probably fascinated by Koizumi's performance but not by his platform, as reported by mass media. This finding confirms the argument that the Japanese people were largely indifferent to politics, even immediately after the 2005 general election (Tsutui et al. 2010).

As stated, Koizumi's popularity is comparable to that of Ronald Reagan (Ohtake 2003, pp. 115–127). Before becoming Governor of California, Reagan was mostly known for his performances as an actor. Our results suggest that Koizumi strategically offered appealing presentations to deliberately attract undecided female voters. Throughout his electoral campaign, Koizumi repeatedly and strongly promoted his political platform by using simple slogans such as "I aim for a small government," "What can be done by the private sector should be done by the private sector," and "I ask whether you agree with the privatization of the postal service" (Ohtake 2003, p. 123). Nonetheless, it was his engaging performance and not his perseverance in his primary policy that probably resulted in his landslide victory. Broadcasting Koizumi's orchestrated political show substantially increased the viewing audience. Inevitably, the TV medium dealt with Koizumi's electoral strategy of "political entertainment" (Ohtake 2003, pp.198–238). Any disadvantages Koizumi might have experienced by not belonging to the mainstream LDP were more than counterbalanced by his unique showmanship and masterful use of mass media. It could be argued that through his experience as a politician, Koizumi naturally learned how effective and significant the superstar effect could be in an election. He exploited TV to purposefully play the dual roles of entertainer and innovator in the political world to overcome the opposition of the traditional political forces linked with special interest groups.

## 5. Conclusions

The objective of this paper was to analyze the role of mass media in the formation of people's consensual support for charismatic political leaders, based on the hypothesis that the superstar effect reported by the literature in several markets may also hold true in the political sector. For this purpose, this study dealt with the case of Koizumi, Japan's longest-serving PM over the past 25 years. To test the hypotheses that media drew voters' attention to Koizumi's charming personal traits rather than to specific policy issues (allowing the PM to fully exploit the "superstar effect"), we empirically assessed whether exposure to TV and newspapers influenced voters' political views.

Based on a survey conducted immediately after Koizumi's landslide electoral win in 2005, a bivariate probit model was used for the estimation. The major findings are as follows. First, results based on combined sample of men and women suggest that the frequencies of exposure to mass media is positively related to the support for Koizumi. On the other hand, they are not significantly correlated with the most distinctive policy he proposed in the electoral campaign. Second, the effect of watching TV on the support for Koizumi is more significant than that of reading newspapers. Third, after dividing the sample into men and women, the impact of watching TV is only observed for females, whereas the influence of reading newspapers is only observed for males.

In contrast to his opponents' case, Koizumi was not supported by the LDP's institutionalized factional system or by special interest groups. Thus, it was necessary for him to use the media to compete with his political rivals. From the observations gathered in this research, we derive the following implications.

Populism has been defined by political theorists as "an appeal to the people against the established structure of power" (Canovan 1999, p. 3) and as "a style of political rhetoric that seeks to mobilize ordinary people against both the established structure of power and the dominant values in society" (Kazin 1995). Our results suggest that Koizumi used the medium of TV as a populist strategy to appeal directly to common people, overcoming the power structure of his own party and challenging the dominant views and practices of the Japanese political establishment. He repeatedly presented himself as the spokesperson of the masses in a struggle with the Japanese elite. From this point of view, Koizumi's political experience notably resembles the rise and success of former Italian PM Silvio Berlusconi (Sabatini 2013).

Koizumi's televised performances played a greater role than his principal policy in drawing electoral support from voters of the opposite sex. This case suggests that TV is a powerful amplifier of political leaders' appealing personal attributes, which may serve a fundamental function in electoral competitions. Our results also seem consistent with the argument that the mix of political news and TV entertainment encourages passive voters to exercise this right. This situation may help outsider candidates to counterbalance the influence of special interest groups (Dyck et al. 2013).

On the other hand, the impact of newspapers is less significant than that of TV. Newspapers can provide detailed information about policy issues, allowing citizens to support specific measures even when they are detrimental to special interest groups. However, exposure to newspapers per se does not seem to counteract the influence of special interest groups; for this purpose, Koizumi's case teaches that the masterful use of TV is a key component of electoral success. Overall, a couple of lessons can be derived. The superstar effect is an important driver of consensus, which may allow celebrities to overcome the opposition of small, special interest groups. At the same time, particular attention should be paid to the danger of populism.

Since the advent of mass media, many charismatic political leaders have effectively used these means of communication to increase voter support. Our results point to TV's particular function in boosting attractive leaders' followers from the opposite sex. Further research is needed to determine whether this argument may be generalized. More specifically, this issue is worth examining by using data from other democracies. The extent to which televised images of a female leader could influence male voters is also a relatively new area of study that merits further exploration.

In this regard, the question arises whether the effect is the same between the sexes. Moreover, this paper has only dealt with the impact of traditional mass media such as TV and newspapers, which do not allow for interactive communication. In modern society, the rising influence of various types of interactive media (such as Twitter and Facebook) on the political arena cannot be ignored (Antoci et al. 2012; Sabatini and Sarracino 2014).Research analyzing the Internet's effects on political opinions is another fairly recent field that warrants in-depth studies.

**References**

Antoci, Angelo, Fabio Sabatini, and Mauro Sodini. 2012. "The Solaria syndrome: Social capital in a growing hyper-technological economy." *Journal of Economic Behavior & Organization* 81(3): 802-814.
Berri, David J., Rob Simmons, Jennifer Van Gilder, and Lisle O'Neill. 2011. "What Does It Mean to Find the Face of the Franchise? Physical Attractiveness and the Evaluation of Athletic Performance." *Economics Letters* 111(3): 200-202.
Biddle, Jeff E. and Daniel S. Hamermesh. 1998. "Beauty, Productivity, and Discrimination: Lawyers' Looks and Lucre." *Journal of Labor Economics* 16(1): 172-201.
Bruni, L., and L. Stanca. 2008. "Watching Alone: Relational Goods, Television and Happiness." *Journal of Economic Behavior & Organization* 65: 506-528.
Canovan, Margaret 1982. "Two Strategies for the Study of Populism. " *Political Studies* 30 (4), 544-552.
Connolly, Sarah, and Shaun P. Hargreaves Heap. 2007. "Cross Country Differences in Trust in Television and the Governance of Public Broadcasters." *Kyklos* 60(1): 3-14.
DellaVigna, Stefano and Ethan Kaplan. 2007. "The Fox News Effect: Media Bias and Voting." *Quarterly Journal of Economics* 122: 1187-1234.
Djankov, Simeon, Caralee McLiesh, Tatiana Nenova, and Andrei Shleifer. 2003. "Who Owns the Media?" *Journal of Law and Economics* 46(2): 341-81.
Doi, Toshihiro, and Takero Ihori. 2009. *The Public Sector in Japan: Past Development and Future Prospects*. Cheltenham: Edward Elgar.
Durante, Ruben, and Brian G., Knight. 2012. "Partisan Control, Media Bias, and Viewers' Responses: Evidence from Berlusconi's Italy." *Journal of European Economic Association* 10(3): 451-481.
Dyck, Alexander, David Moss, and Luigi Zingales. 2013. "Media versus Special Interests." *Journal of Law and Economics* 56: 521-553.
Enikolopov, Ruben, Maria Petrova, and Ekaterina Zhuravskaya. 2011. "Media and Political Persuasion: Evidence from Russia." *American Economic Review* 101(7): 3253-3285.
Ehrmann, Thomas, Brinja Meiseberg, Christian Ritz 2009. Superstar Effects in Deluxe Gastronomy – An Empirical Analysis of Value Creation in German Quality Restaurants. *Kyklos* 62 (4), 526-541.
Frey, Bruno S., Christine Benesch, Alois Stutzer. 2007. "Does Watching TV Make us Happy?" *Journal of Economic Psychology* 28: 283-313.
Gentzkow, Matthew. 2006. "Television and Voter Turnout." *Quarterly Journal of Economics* 121(3):931-72.

Gentzkow, Matthew, and Jesse M. Shapiro. 2004. "Media, Education and Anti-Americanism in the Muslim World." *Journal of Economic Perspectives* 18(3): 117-33.
Geroge, Lisa and Joel Waldfogel. 2006. "The New York Times and the Market for Local Newspapers." *American Economic Review* 96: 435-447.
Gerber, Alan, Dean Karlan, and Daniel Bergan. 2009. "Does the Media Matter? A Field Experiment Measuring the Effect of Newspapers on Voting Behavior and Political Opinions." *American Economic Journal: Applied Economics* 1(2): 35-52.
Grabe, Maria Elizabeth, Erik Page Bucy 2009. *Image Bite Politics: News and the Visual Framing of Elections*. Oxford University Press.
Greene, William. 2008. *Econometric Analysis* (6 eds). London: Prentice-Hall.
Hamermesh, Daniel. S. and Jeff E. Biddle. 1994. "Beauty and the Labor Market." *American Economic Review* 84(5): 1174-1194.
Hayashi, Chikio, Masafumi Sakuraba. 2002. *Data suggest Japanese potential* (Suuji ga Akasu Nihonjin no Senzairyoku). Tokyo: Kodan-sha.
Imai, Masami 2009. "Ideologies, Vested Interest Groups, and Postal Saving Privatization in Japan." *Public Choice* 138: 137-160.
Katz, Richard. 2001. "Koizumi is blowing it." *International Economy* 15(5): 38-39, 51.
Kazin, Michael. 1995. *The Populist Persuasion: An American History*. New York. Basic Books.
Köllner, Patrick. 2006. "The Liberal Democratic Party at 50: Sources of Dominance and Changes in the Koizumi Era." *Social Science Japan Journal* 9(2): 43-57.
Krauss, Ellis S. 2002. "The Media's Role in a Changing Japanese Electorate." 6-12 in *Undercurrents in Japanese Politics*, Asia Program Special Report, 2nd ed., Woodrow Wilson International Center for Scholars.
Krauss, Ellis S. and Benjamin Nyblade. 2005. "Presidentialization" in Japan? The Prime Minister, Media and Elections in Japan." *British Journal of Political Science* 35: 357-368.
Lawson, C, and J.A. McCann. 2005. "Television News, Mexico's 2000 Elections, and Media Effects in Emerging Democracies." *British Journal of Political Science* 35(1): 1-30.
Lindgren, Petter. 2012. The Era of Koizumi's Right-Wing Populism. University of Oslo, Department of Cultural Studies and Oriental Languages, mimeo.
Lucifora, Claudio, Rob Simmons. 2003. "Superstar Effects in Sport Evidence From Italian Soccer". *Journal of Sports Economics* 4 (1), 35-55.
Miyake, Ichiro. 1989. *Tohyo kodo* (in Japanese) *(The Voting Behavior)*. Tokyo: Tokyo University Press.
Oberholzer-Gee, Felix and Joel Waldfogel. 2009. "Media Markets and Localism: Does Local News en Espanol Boost Hispanic Voter Turnout?" *American Economic Review* 99(5):2120–28.
Ohtake, H. 2003. *Populism in Japan: Anticipation and Disillusion about Politics* (Nihon-gata Populism: Seiji-eno Kitaki to Genmetsu in Japanese). Tokyo: Chuokoron-sha.
Olken, Benjamin. 2009. "Do Television and Radio Destroy Social Capital? Evidence from Indonesian Villages." *American Economic Journal: Applied Economics* 1(4): 1-33.
Petrova, Maria. 2008. Inequality and Media Capture. *Journal of Public Economics*, 92(1–2):1 83-212.
Rosen, Sherwin 1981. "The Economics of Superstars." *American Economic Review* 71: 845-858.
Sabatini, Fabio. 2012. "Who Trusts Berlusconi? An Econometric Analysis of the Role of Television in the Political Arena." *Kyklos* 65(1): 110-130.
Sabatini, Fabio and Francesco Sarracino. 2014. Will Facebook Save or Destroy Social Capital? An Empirical Investigation into the Effect of Online Interactions on Trust and Networks. EERI Research Paper Series EERI RP 2014/02.
Sakakibara, Eisuke. 2003. *Structural Reform in Japan: Breaking the Iron Triangle*. Washington, D.C.: Brookings Institution Press.
Stolberg, S.G. and S. Dewan. 2006. "In Memphis, Two Heads of Government Visit the Home of Rock 'N' Roll Royalty. *New York Times,* July 1.

Strom, Stephanie. 2001. Plain-spoken leader; Junichiro Koizumi. *New York Times,* April 25.

Strömberg, David. 2004. "Radio's Impact on Public Spending." *Quarterly Journal of Economics* 119(1): 189-221.

Tsutsui, Yoshiro, Miles Kimball, and Fumio Ohtake. 2010. "Koizumi Carried the Day: Did the Japanese Election Results Make People Happy and Unhappy?" *European Journal of Political Economy* 26(1):12-24.

Van Gelder, Lawrence. 2005. Follow the leader. *New York Times,* May 30, 2005.

White, Stephen, Sarah Oates, and Ian McAllister. 2005. "Media Effects and the Russian Elections, 1999–2000." *British Journal of Political Science* 35(2): 191-208.

Yamamura, Eiji, and Haruo Kondoh. 2013. "Government Transparency and Expenditure in the Rent-seeking Industry: The Case of Japan for 1998–2004. " *Contemporary Economic Policy* 31(3): 635-647.

**Tables and figures**

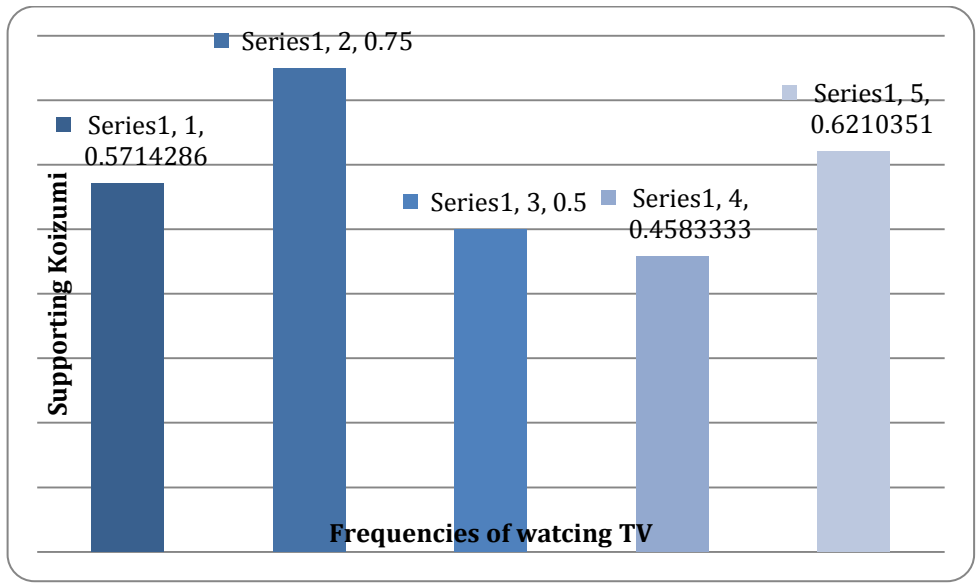

(a) Men

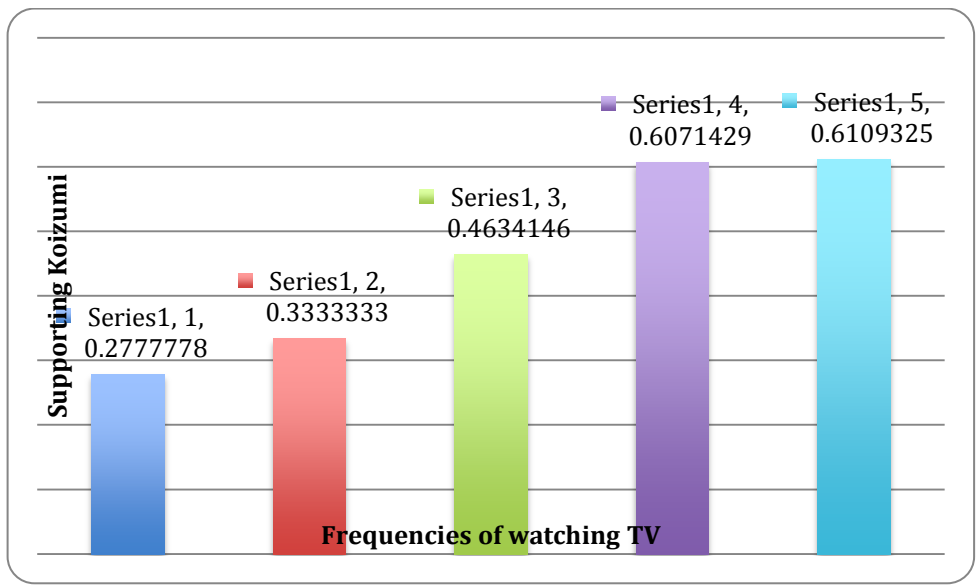

(b) Women

Figure 1. Supporting for koizumi regime

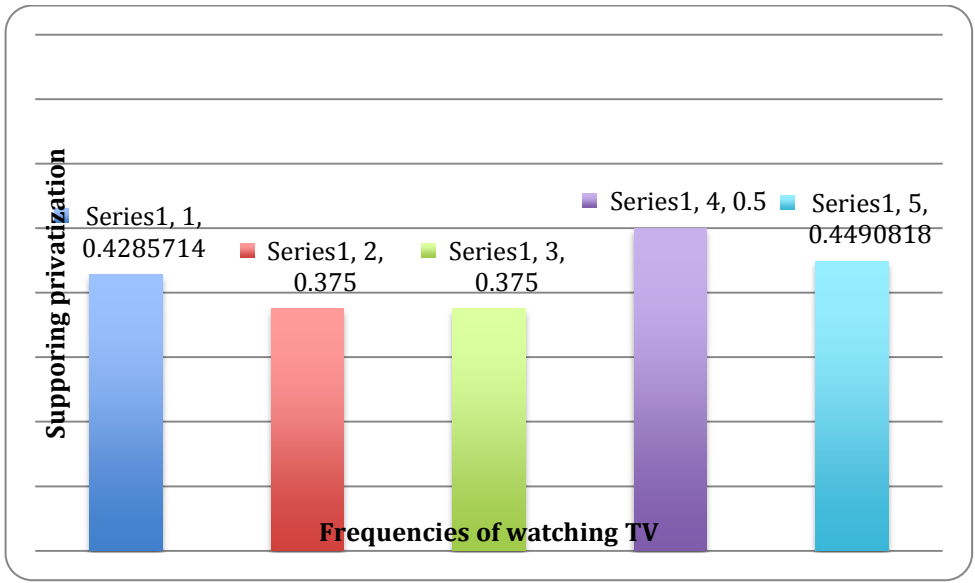

(a) Men

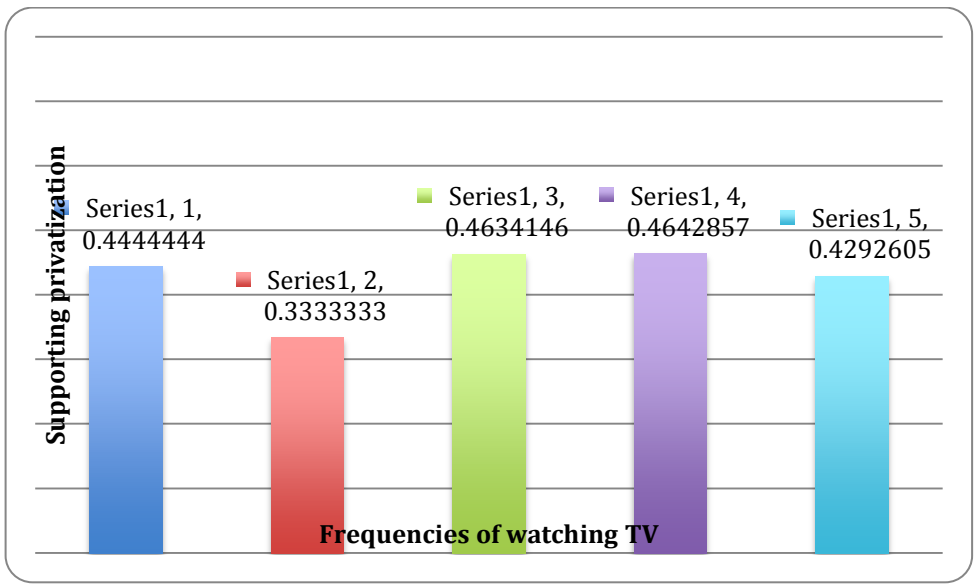

(b) Women

Figure 2. Supporting privatization of postal service.

Table 1
Duration of each government (from 1989 to 2012)

| Government | Months | Highest support rate (%) | Lowest support rate (%) |
|---|---|---|---|
| Uno | 2 | 28 | 28 |
| Kaifu | 27 | 56 | 35 |
| Miyazawa | 21 | 54 | 20 |
| Hosokawa | 8 | 71 | 57 |
| Hata | 2 | 47 | 47 |
| Murayama | 19 | 42 | 33 |
| Hashimoto | 30 | 53 | 31 |
| Obuchi | 21 | 49 | 23 |
| Mori | 12 | 39 | 7 |
| **Koizumi** | **60** | **85** | **39** |
| Abe (First government) | 12 | 65 | 29 |
| Fukuda | 12 | 58 | 20 |
| Aso | 12 | 49 | 15 |
| Hatoyama | 9 | 72 | 21 |
| Kan | 15 | 65 | 16 |
| Noda | 15 | 60 | 20 |
| Average | 17.3 | | |

Note: From Uno to Obuchi administrations, the data is collected from Hayamashi and Sakuraba (2002). After Mori administration, the data is gathered from annual investigation about political view cudnucated byd Nippon Hoso-kyokai.
Source: http://www2.ttcn.ne.jp/honkawa/5236a.html (accessed on April 26, 2014).

Table 2
Variable definitions and descriptive statistics

| Variables | Definition | Mean |
|---|---|---|
| KOIZUMI | Value is 1 if respondent supports Koizumi, 0 otherwise. | 0.62 |
| PRIVAT | Value is 1 if respondent supports the privatization of postal services, otherwise 0. | 0.42 |
| TV | Frequency of watching TV. 1 (not at all)–5 (every day) | 4.75 |
| NEWS | Frequency of reading newspaper. 1 (not at all)–5 (every day) | 4.38 |
| INCOM | Household income (million yen) | 5.22 |
| EDU | Years of schooling | 12.5 |
| AGE | Age | 54.7 |
| SPOUS | Value is 1 if respondent has a spouse, otherwise 0. | 0.80 |
| CHILD | Value is 1 if respondent has a child, otherwise 0. | 0.55 |
| UNEMP | Value is 1 if respondent is unemployed, otherwise 0. | 0.19 |
| KNOW | Value is 1 if respondent knows that Yohei Kohno is the Speaker of the House of Representatives in 2005, otherwise 0. | 0.35 |
| GOVPT1 | Value is 1 if respondent supports the government party (the Liberal Democratic Party), otherwise 0. | 0.39 |
| GOVPT2 | Value is 1 if respondent supports the government party (the Komei Party), otherwise 0. | 0.03 |
| VILLAG | Value is 1 if respondent lives in town/village, otherwise 0. | 0.17 |
| MAN | Value is 1 if man, 0 if woman. | 0.53 |

Table 3
Determinants of supporting the Koizumi regime and the postal reformation
(Bivariate probit model): full sample

| | (1) KOIZUMI | (2) PRIVAT | (3) KOIZUMI | (4) PRIVAT | (5) KOIZUMI | (6) PRIVAT |
|---|---|---|---|---|---|---|
| TV | 0.03* | −0.01 | 0.04** | −0.002 | | |
| | (1.70) | (−0.23) | (2.49) | (−0.10) | | |
| NEWS | 0.01 | 0.01 | | | 0.02** | 0.004 |
| | (1.45) | (0.40) | | | (2.29) | (0.31) |
| Ln(INCOM) | 0.07*** | 0.001 | 0.07*** | 0.002 | 0.06** | 0.001 |
| | (2.67) | (0.06) | (2.78) | (0.09) | (2.59) | (0.07) |
| Ln(EDU) | −0.04 | 0.20** | −0.03 | 0.20** | −0.03 | 0.20** |
| | (−0.61) | (2.41) | (−0.49) | (2.47) | (−0.54) | (2.44) |
| Ln(AGE) | −0.14** | 0.001 | −0.13** | 0.01 | −0.14** | 0.001 |
| | (−2.49) | (0.02) | (−2.34) | (0.15) | (−2.51) | (0.01) |
| SPOUS | −0.01 | 0.02 | 0.002 | 0.02 | 0.001 | 0.02 |
| | (−0.17) | (0.50) | (0.01) | (0.53) | (0.04) | (0.46) |
| CHILD | −0.03 | 0.01 | −0.03 | 0.01 | −0.03 | 0.01 |
| | (−1.09) | (0.21) | (−0.99) | (0.23) | (−1.18) | (0.27) |
| UNEMP | 0.05 | 0.01 | 0.05 | 0.01 | 0.04 | 0.01 |
| | (1.20) | (0.14) | (1.24) | (0.12) | (1.07) | (0.18) |
| KNOW | −0.01 | −0.03 | −0.004 | −0.03 | −0.01 | −0.03 |
| | (−0.29) | (−1.01) | (−0.14) | (−0.97) | (−0.31) | (−0.98) |
| GOVPT1 | 0.44*** | 0.17*** | 0.44*** | 0.17*** | 0.44*** | 0.17*** |
| | (19.9) | (5.19) | (19.7) | (5.20) | (20.1) | (5.15) |
| GOVPT2 | 0.24*** | 0.07 | 0.25*** | 0.07 | 0.25*** | 0.07 |
| | (3.29) | (0.79) | (3.37) | (0.81) | (3.34) | (0.79) |
| VILLAG | −0.10*** | −0.04 | −0.10*** | −0.04 | −0.10*** | −0.04 |
| | (−2.68) | (−0.95) | (−2.67) | (−0.93) | (−2.67) | (−0.93) |
| MAN | −0.04 | −0.01 | −0.04 | −0.003 | −0.04 | −0.004 |
| | (−1.28) | (−0.14) | (−1.26) | (−0.14) | (−1.25) | (−0.12) |
| Wald chi-sq | 208 | | 205 | | 202 | |
| Sample size | 858 | | 858 | | 858 | |

Numbers without parentheses show marginal effects. Numbers in parentheses are z-statistics based on robust-standard errors. *, ** and *** indicate significance at 10, 5 and 1% levels, respectively. The constant is included but its result is not reported.

Table 4
Determinants of support for Koizumi and the postal privatization
(Bivariate probit model): sample of men

|        | (1) KOIZUMI | (2) PRIVAT | (3) KOIZUMI | (4) PRIVAT | (5) KOIZUMI | (6) PRIVAT |
|--------|-------------|------------|-------------|------------|-------------|------------|
| TV     | 0.002       | 0.002      | 0.04        | 0.01       |             |            |
|        | (0.88)      | (0.07)     | (1.57)      | (0.39)     |             |            |
| NEWS   | 0.02        | 0.01       |             |            | 0.03*       | 0.01       |
|        | (1.44)      | (0.78)     |             |            | (1.91)      | (0.88)     |
| Ln(INCOM) | 0.06*    | 0.04       | 0.06*       | 0.04       | 0.06*       | 0.04       |
|        | (1.73)      | (1.00)     | (1.85)      | (1.09)     | (1.71)      | (1.00)     |
| Ln(EDU) | −0.01      | 0.22**     | 0.001       | 0.23**     | −0.004      | 0.22**     |
|        | (−0.05)     | (2.04)     | (0.01)      | (2.09)     | (−0.05)     | (2.04)     |
| Ln(AGE) | −0.11      | 0.002      | −0.10       | 0.01       | −0.11       | 0.002      |
|        | (−1.24)     | (0.02)     | (−1.08)     | (0.10)     | (−1.22)     | (0.03)     |
| SPOUS  | −0.08       | 0.03       | −0.07       | 0.04       | −0.07       | 0.03       |
|        | (−1.34)     | (0.53)     | (−1.15)     | (0.64)     | (−1.18)     | (0.54)     |
| CHILD  | −0.01       | 0.01       | −0.01       | 0.02       | −0.01       | 0.01       |
|        | (−0.34)     | (0.39)     | (−0.29)     | (0.42)     | (−0.36)     | (0.39)     |
| UNEMP  | 0.05        | 0.002      | 0.06        | 0.01       | 0.05        | 0.001      |
|        | (0.95)      | (0.03)     | (1.02)      | (0.09)     | (0.90)      | (0.03)     |
| KNOW   | 0.02        | −0.08*     | 0.02        | −0.08*     | 0.02        | −0.08*     |
|        | (0.49)      | (−1.83)    | (0.62)      | (−1.76)    | (0.40)      | (−1.85)    |
| GOVPT1 | 0.41***     | 0.19***    | 0.41***     | 0.18***    | 0.41***     | 0.19***    |
|        | (13.7)      | (4.25)     | (13.6)      | (4.21)     | (13.7)      | (4.25)     |
| GOVPT2 | 0.37***     | 0.13       | 0.37***     | 0.14       | 0.37***     | 0.13       |
|        | (2.89)      | (1.08)     | (2.94)      | (1.11)     | (2.87)      | (1.07)     |
| VILLAG | −0.12**     | −0.06      | −0.12**     | −0.05      | −0.12**     | −0.06      |
|        | (−2.59)     | (−1.05)    | (−2.54)     | (−1.00)    | (−2.62)     | (−1.05)    |
| Wald chi-sq | 116    |            | 114         |            | 115         |            |
| Sample size | 455    |            | 455         |            | 455         |            |

Numbers without parentheses show marginal effects. Numbers in parentheses are z-statistics based on the robust-standard errors. *, ** and *** indicate significance at 10, 5 and 1% levels, respectively. The constant is included but its result is not reported.

Table 5
Determinants of support for Koizumi and the postal privatization
(Bivariate probit model): sample of women

|  | (1) KOIZUMI | (2) PRIVAT | (3) KOIZUMI | (4) PRIVAT | (5) KOIZUMI | (6) PRIVAT |
|---|---|---|---|---|---|---|
| TV | 0.04* | −0.01 | 0.05** | −0.01 | | |
|  | (1.69) | (−0.54) | (2.16) | (−0.68) | | |
| NEWS | 0.01 | −0.004 | | | 0.02 | −0.01 |
|  | (0.82) | (−0.21) | | | (1.51) | (−0.46) |
| Ln(INCOM) | 0.09** | −0.04 | 0.09** | −0.04 | 0.09** | −0.04 |
|  | (2.42) | (−0.96) | (2.47) | (−0.97) | (2.30) | (−0.96) |
| Ln(EDU) | −0.10 | 0.20 | −0.09 | 0.20 | −0.09 | 0.21* |
|  | (−1.01) | (1.64) | (−0.93) | (1.60) | (−0.93) | (1.68) |
| Ln(AGE) | −0.16** | −0.04 | −0.15** | −0.04 | −0.16** | −0.04 |
|  | (−2.06) | (−0.48) | (−2.04) | (−0.49) | (−2.11) | (−0.45) |
| SPOUS | 0.03 | 0.02 | 0.04 | 0.02 | 0.03 | 0.02 |
|  | (0.68) | (0.39) | (0.78) | (0.36) | (0.72) | (0.37) |
| CHILD | −0.04 | −0.01 | −0.04 | −0.01 | −0.05 | −0.01 |
|  | (−1.03) | (−0.28) | (−0.95) | (−0.31) | (−1.16) | (−0.19) |
| UNEMP | 0.03 | 0.07 | 0.03 | 0.07 | 0.01 | 0.08 |
|  | (0.42) | (0.85) | (0.43) | (0.84) | (0.16) | (0.95) |
| KNOW | −0.07 | 0.06 | −0.06 | 0.06 | −0.06 | 0.06 |
|  | (−1.31) | (1.04) | (−1.21) | (1.02) | (−1.17) | (1.04) |
| GOVPT1 | 0.49*** | 0.15*** | 0.49*** | 0.15*** | 0.50*** | 0.15*** |
|  | (13.9) | (3.13) | (13.8) | (3.15) | (14.1) | (3.05) |
| GOVPT2 | 0.16* | 0.02 | 0.17* | 0.02 | 0.17* | 0.02 |
|  | (1.87) | (0.20) | (1.92) | (0.21) | (1.91) | (0.21) |
| VILLAG | −0.07 | −0.01 | −0.07 | −0.01 | −0.07 | −0.01 |
|  | (−1.38) | (−0.13) | (−1.41) | (−0.12) | (−1.26) | (−0.13) |
| Wald chi-sq | 117 | | 114 | | -466 | |
| Sample size | 404 | | 404 | | 404 | |

Numbers without parentheses show marginal effects. Numbers in parentheses are z-statistics based on the robust-standard errors. *, ** and *** indicate significance at 10, 5 and 1% levels, respectively. The constant is included but its result is not reported.